\documentclass[onecolumn,showpacs,showkeys,preprintnumbers,superscriptaddress,amsmath,amssymb]{revtex4}
\usepackage[dvips]{graphicx}
\usepackage{latexsym,amssymb,amsmath,epsfig,bm,times,psfrag}
\usepackage{color}
\usepackage[bookmarksnumbered,bookmarksopen,colorlinks,citecolor=red,linkcolor=blue]{hyperref}
\usepackage{epstopdf}
\newcommand{\Fig}[1]{Fig.~\ref{#1}}
\newcommand{\eq}[1]{Eq.~(\ref{#1})}
\newcommand{\sect}[1]{Sec.~\ref{#1}}

\renewcommand{\part}{{\rm part}}

\newcommand{\be}{\begin{equation}}
\newcommand{\ee}{\end{equation}}
\newcommand{\bear}{\begin{eqnarray}}
\newcommand{\eear}{\end{eqnarray}}
\newcommand{\ba}{\begin{array}}
\newcommand{\ea}{\end{array}}

\begin{document}


\title{Volume fluctuations affect on transport and thermodynamic coefficients in p-Pb systems}

\author{De-Xian Wei}
\email{dexianwei@gxust.edu.cn}
\affiliation{School of Science, Guangxi University of Science and Technology, Liuzhou, 545006, China}
\date{\today}

\begin{abstract}
We use A Multi-Phase Transport (AMPT) model to simulate the event-by-event p-Pb collisions at $\sqrt{s_{NN}}$=5.02 TeV.
To study the space-time volume fluctuations, two different definitions of radius have been introduced in the calculation: 
One is weigh of event-by-event charged multiplicity (noted as set I), and the other one is weigh of energy density in a single event (notes as set II).
Based on calculations of sets I and II, the transport and thermodynamic coefficients, such as the speed of sound squared, shear viscosity, bulk viscosity, and mean free path are both dependent on the radius and local temperature.
By comparing results of sets I and II, we found that these transport and thermodynamic coefficients differed significantly in the results.
These results imply that the transport and thermodynamic properties of the medium in small collisions significantly depend on the space-time volume fluctuations.
\end{abstract}



\maketitle


\section{Introduction}
\label{sec:introdution}

Heavy-ion collision experiments at the Relativistic Heavy Ion Collider (RHIC) and Large Hadron Collider (LHC) have generated the hot and dense state of strongly interacting medium known as the Quark Gluon Plasma (QGP) in large colliding systems such as Au-Au~\cite{Arsene:2005qgp,Back:2005tpp,Adams:2005eat,Adcox:2005fod} and Pb-Pb~\cite{Muller:2012frf}.
One of the signatures of QGP is the collective flow, which can be described well by hydrodynamics model~\cite{Heinz:2013cfa}.
Similar observations of collective flow have been found in p+Pb collisions at LHC~\cite{Khachatryan:2015efc} and in p+Au, d+Au, and $^{3}$He+Au collisions at RHIC~\cite{Aidala:2019coq,Nagle:2018ssc}.
The similar collective behavior observed in p+A and A+A systems can be explained if a strongly coupled QGP is formed in all these systems~\cite{Weller:2017oft}. However, no evidence has been found in the formation of a QGP droplet in small collisions of hard probes, such as high $p_{T}$ jet quenching~\cite{Nagle:2018ssc,Khachatryan:2017cnm,Oetringhaus:2024ado}.

If QGP droplet is confirmed in small collisions, a nature unavoidable question:~What is the smallest size of a droplet of strongly couple hydrodynamic like matter? To answer this question, one needs to carry out the crucial of effective framework with the initial conditions and transport properties to describe the system in evolution.
A number of phenomenological studies on the transport properties of QGP (noted specific viscosities) in relativistic heavy-ion collisions have been studied over past years. The calculations on theory using a variety of approaches include the Yang-Mills theory~\cite{Haas:2014gsf}, non-conformal holography~\cite{Grefa:2022tco}, excluded-volume hadron resonance gas (EHRG) model~\cite{McLaughlin:2022bat}, Polyakov-Nambu-Jona-Lasinio (PNJL) models~\cite{Islam:2021iod}, Color String Percolation Model (CSPM)~\cite{Sahu:2021tat}, and dynamical quasi-particle model (DQPM)~\cite{Soloveva:2020tcf}, etc. While the fermion sign problem on strongly-coupled nature of QGP~\cite{Troyer:2005cca,Altenkort:2023vop}, one cannot currently calculate these transport coefficients from the first principle.
The impact of finite volume on the transport and thermodynamic properties have also been studied~\cite{Saha:2018tci, Chahal:2023eof,Pal:2024eof,Shaikh:2024qpd,Sharma:2024ssd}. For different sizes of systems, the different values of the thermodynamic variables imply that differential pre-hydrodynamic phase generated in these systems~\cite{Nijs:2021tmd,Nijs:2021bao,Sharma:2024ssd}.
And ref.~\cite{Sun:2024veo} also pointed out that a sizable volume effect can be induced by the initial-state fluctuations in the TRENTo model. While the initial-state fluctuations in small collisions are still not fully understood.

In recent years, data-driven techniques-machine learning and Bayesian analysis methods have been developed and currently are successful in extracting the transport coefficients from data from small to large systems~\cite{Nijs:2021bao,Bernhard:2019beo,Moreland:2020bco,Everett:2021mbc,Parkkila:2021beo, Everett:2021pco,Heffernan:2024bqo}(also seen in a review\cite{Achenbach:2024tpa}). While these analysis depend on the values assumed for many parameters, so reliable determination of the realtime QGP properties requires a systematic examination.

The transport and thermodynamic properties in small systems remain open questions. The motivation of this work: to study volume fluctuations in event-by-event p-Pb collisions employing by A Multi-Phase Transport (AMPT) model~\cite{Lin:2004amt}, and extracted the effective transport and thermodynamic coefficients. This work is organized as follows:
\sect{sec:method} using the kinetic theory approach to extract the transport and thermodynamic coefficients. These coefficients are calculated by different radius sets and volume cut-off methods. The numerical results are presented in \sect{sec:result}.
Summary are given in \sect{sec:sum}.
We use natural unit $k_B=c=\hbar=1$.

\section{Materials and Methods}
\label{sec:method}

In a many-body system, one describe the space-time information of particles is the distribution function $f[r(t),p(t)]$. The distribution function can be defined by~\cite{Bertsch:1988agt}
\begin{eqnarray}\label{disfun:01}
f[r(t),p(t)] &=& \frac{1}{N}\sum_{i=1}^{N}\delta^{(3)}[r(t)-r_{i}(t)]\delta^{(3)}[p(t)-p_{i}(t)]
\end{eqnarray}
where $N$ is the total number of test particles in a single event, $r_{i}$ and $p_{i}$ are the time-dependent coordinate and momentum of the $i$th source particle in a single event, $r$ and $p$ are the time-dependent coordinate and momentum of the observation point, and the summation is over all the particles.

To obtain the effective transport and thermodynamic properties, one defines the energy-momentum tensor from the distribution function in kinetic theory
\begin{eqnarray}\label{emten:01}
T^{\mu\nu}(t,r) &=& \int\frac{d^{3}p}{(2\pi)^{3}}\frac{p^{\mu}p^{\nu}}{p^{0}}f[r(t),p(t)]
\end{eqnarray}

The pressure coefficient of a thermodynamic system is given by
\begin{eqnarray}\label{tans:01}
\beta &=& \frac{1}{P}\frac{\partial P}{\partial T}
\end{eqnarray}
where the temperature $T=\frac{1}{3}\langle p_{T}\rangle$, the energy density and pressure density are obtained from \eq{emten:01}, i.g., $\varepsilon=T^{00},~P=\sqrt{T^{11}+T^{22}+T^{33}}$. The pressure coefficient is a quantity that the system deviates from ideal medium. In the case $\beta=\frac{1}{T}$ means that the medium is close to ideal matter, while $\beta\neq\frac{1}{T}$ means that the medium is deviates from ideal matter due to the viscosity.

It is crucial to investigate transport coefficients since they shed light on the genesis of the medium. The transport coefficients are defined as coefficients of the space-space component of a deviation of the energy momentum tensor from equilibrium. In a medium composed of fermions with the momentum distribution function $f\equiv f[r(t),p(t)]$, the transport coefficients of the medium, e.g., shear viscosity ratio and bulk viscosity ratio can be obtained by the kinetic theory~\cite{Bozek:2010bas,Zanna:2013rvh,Garcia:2023von}
\begin{eqnarray}\label{tans:02}
\eta/s &=& \frac{1}{15(\varepsilon+P)}\int\frac{d^{3}p}{(2\pi)^{3}}\frac{p^{4}}{p^{0}}[f(1-f)]
\end{eqnarray}
\begin{eqnarray}\label{tans:03}
\zeta/s &=& 15 \eta/s\left[\frac{1}{3}-c_{s}^{2}\right]^{2}
= \frac{1}{(\varepsilon+P)}\left[\frac{1}{3}-c_{s}^{2}\right]^{2}  \int\frac{d^{3}p}{(2\pi)^{3}}\frac{p^{4}}{p^{0}}[f(1-f)]
\end{eqnarray}
where the entropy density $s=\frac{(\varepsilon+P)}{T}$, and the speed of sound squared, which is defined as the ratio of the pressure to the energy density, $c_{s}^{2}=\frac{\partial P}{\partial \varepsilon}$.

The mean free path is an important thermodynamical quantity which is defined as the average distance traveled by a particle between two successive collisions. The events averaged mean free path is defined as~\cite{Sahu:2021tat}
\begin{eqnarray}\label{tans:04}
\lambda &\simeq& \frac{5}{T}\eta/s =\frac{1}{3(\varepsilon+P)T}\int\frac{d^{3}p}{(2\pi)^{3}}\frac{p^{4}}{p^{0}}[f(1-f)]
\end{eqnarray}
 
In relativistic collision system, the definition of the radius is not unique. We define two types of radius, $R_{I}$ and $R_{II}$, where $R_{I}$ is weigh of event-by-event charged multiplicity, as~\cite{Acharya:2024cpa}
\begin{eqnarray}\label{rad:01}
R_{I}(t) &=& \frac{N_{ch}(t)}{\langle N_{ch}(t)\rangle}
\end{eqnarray}
where $N_{ch}$ is charged multiplicity for a single event, $t$ is the evolution time, $\langle\cdot\rangle$ is the events averaged charged multiplicity. The charged multiplicity fluctuating in event-by-event implies that volume is fluctuating for event-by-event in \eq{rad:01}. And $R_{II}$ is weigh of the energy density in a single event, as~\cite{Sievert:2019clh}
\begin{eqnarray}\label{rad:02}
R_{II}(t) &=& \left[\frac{\int r^{2}(t)\varepsilon(r(t),\phi(t))rdrd\phi}{\int \varepsilon(r(t),\phi(t))rdrd\phi}\right]^{\frac{1}{2}}
\end{eqnarray}
where $\varepsilon(r,\phi)$ is the local energy density in a single event.
Due to the fluctuations of initial geometry, the fluctuating local energy density is corresponding to a fluctuating volume in a single event. As a consequence, the volume fluctuations in evolution are major influenced by initial fluctuations. By fixing radius of a system at the time stamp, the transport and thermodynamic coefficients can perform statistics, used by \eq{rad:01} and \eq{rad:02}, respectively. Of course, one must be careful in assessing the fluctuations versus
the non-flow effects in small systems. Note that it cannot suppress the non-flow effects (e.g., partons in hard scattering) in kinetic theory. Such residual non-flow effects may also affect on the transport and thermodynamic coefficients.
  
In this work, we use AMPT model to simulate the event-by-event p-Pb collisions. The AMPT model includes the transport process that simulates parton interactions and has a natural shear viscosity that is analytically or numerically calculable, dependent on the scattering cross section, in kinetic theory such as Israel-Stewart or Chapman-Enskog methods~\cite{MacKay:2022tsv,Zhang:2023svt}. In this work, we consider the partons anisotropic scattering system evolution in a finite volume with fluctuations. Therefore, we re-extract the effective transport and thermodynamic coefficients employed by kinetic theory, as Eqs.~(\ref{tans:01})-(\ref{tans:04}). These coefficients are calculated using the volume cut-off methods.

In this work, the transport and thermodynamic coefficients are investigated in p-Pb (b=0 fm) collisions at $\sqrt{s_{NN}}$ = 5.02 TeV. These coefficients are re-extracted in partons evolution stage. 
Note that the impact parameters (b=0 fm) are controlled by the transverse distances of the overlap in initial collision space, not the final charged participant multiplicity, $M$. As a consequence, it includes a wide multiplicity range, from low $M$ to large ones. The transverse overlap in initial collision geometry is fluctuating in event-by-event basis.
This work takes the collisions probabilities parameters in AMPT with $a=0.5$, $b=0.9$ GeV$^{-2}$, $\alpha_{s}$ =0.33 and $\mu$ = 3.2 fm$^{-1}$ for p-Pb systems are consistent with our previous work~\cite{Wei:2020rli}. More thorough details of the AMPT model can be found in Ref.~\cite{Lin:2004amt}. Note also that in this work, we do not attempt to compare simulations with experimental data, but rather to explore how the finite volume and its fluctuations influence on the transport and thermodynamic coefficients in event-by-event small collisions.

\section{Results}
\label{sec:result}

\begin{figure*}[tp]
\begin{center}
\includegraphics[width=0.40\textwidth]{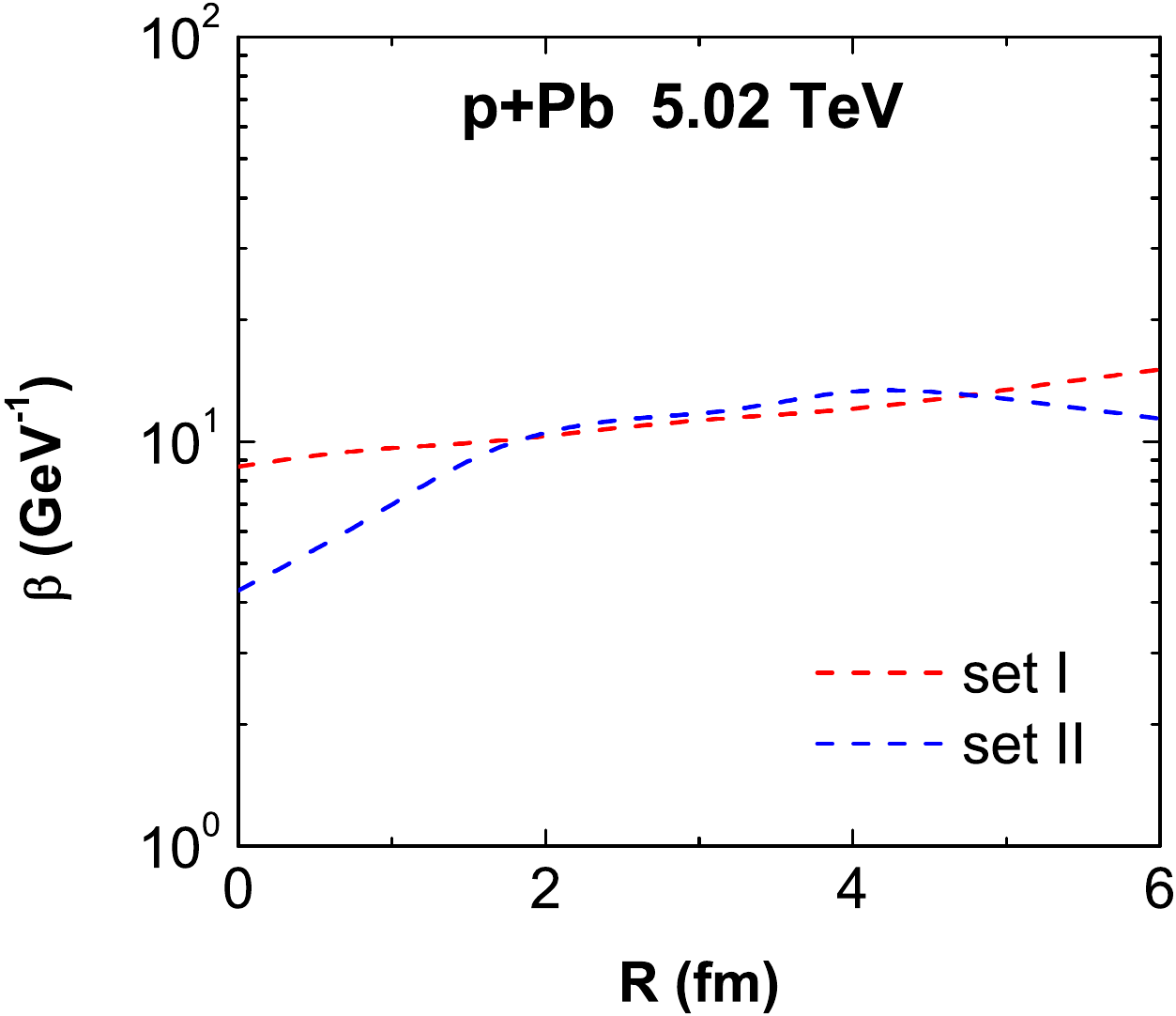}
\caption{(Color online)
The pressure coefficient $\beta$ as a function of radius.
}
\label{fig1}
\end{center}
\end{figure*}

To investigate the fluctuating properties in small collisions, we show the events averaged pressure coefficient in \Fig{fig1}. As illustrated in \Fig{fig1}, the events averaged pressure coefficient $\beta$ as a function of radius.
By comparing the results of sets I and II, the behavior of $\beta$ linear dependent on radius are significantly similar, except for a smaller radius ($R< 2$ fm) and larger radius ($R> 5$ fm) in set II. For set I, the radius depends on the event multiplicity, and naturally, the events averaged pressure coefficient exhibits event-by-event fluctuations; For set II, the radius depends on the local energy density of a single event, as a result, the pressure coefficient reflects the fluctuating nature of the local energy density in an event. These fluctuations, in fact, correspond to volume fluctuations. It can be concluded that the medium exhibits different ways of deviating from ideal matter due to different volume fluctuations.

\begin{figure*}[tp] 
\begin{center}
\includegraphics[width=0.30\textwidth]{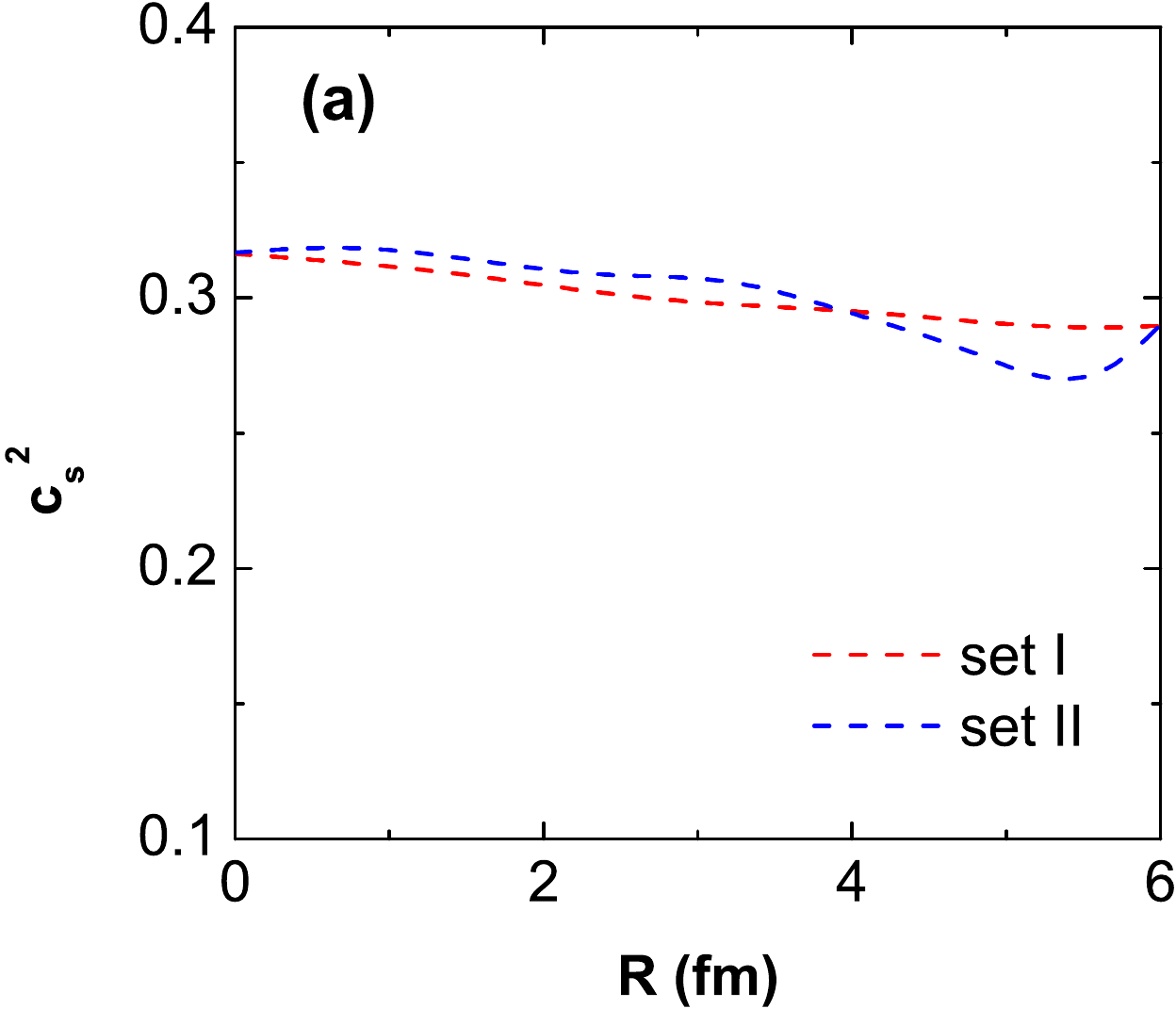}
\hspace{0.10cm}
\includegraphics[width=0.30\textwidth]{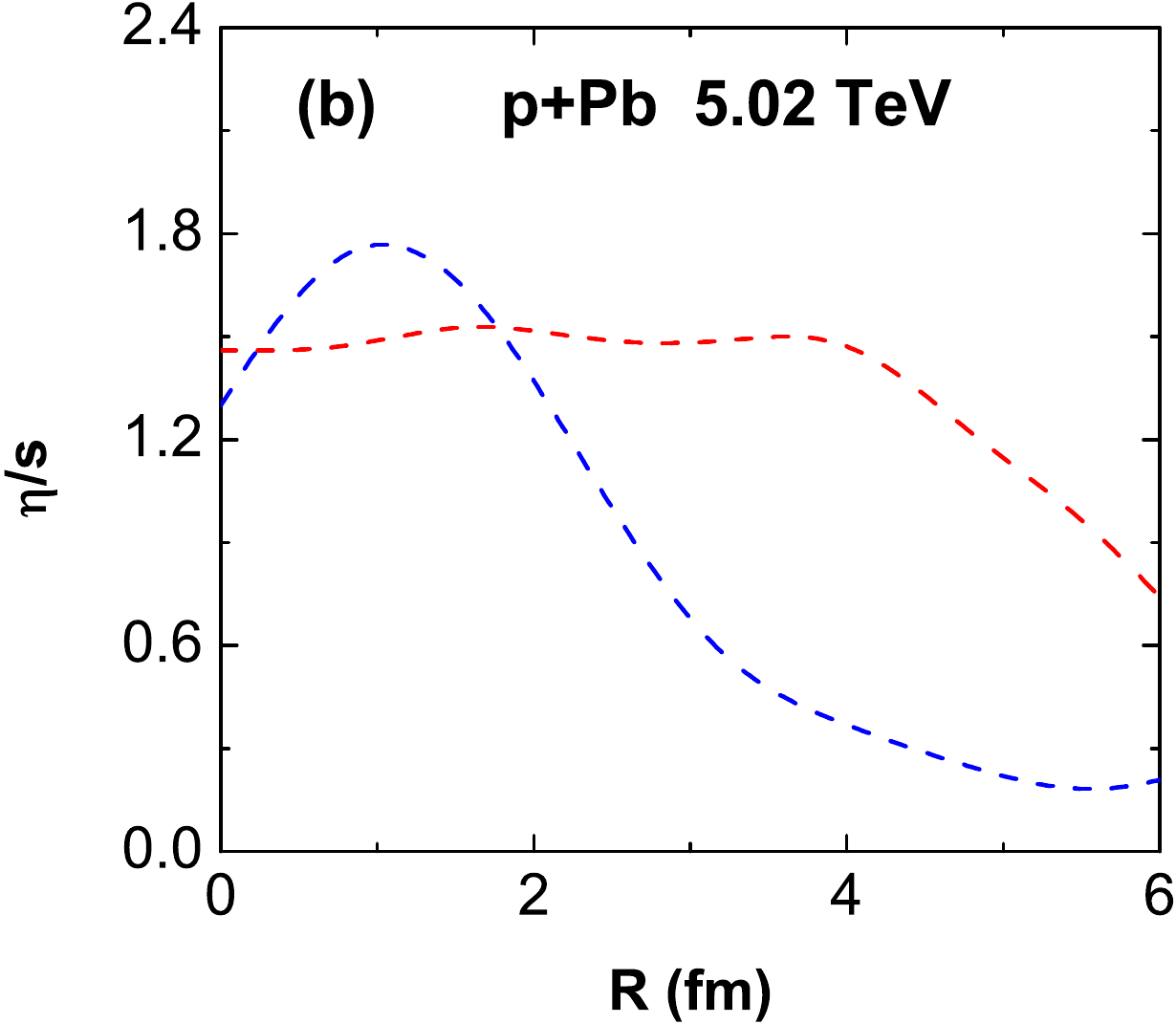}
\hspace{0.10cm}
\includegraphics[width=0.30\textwidth]{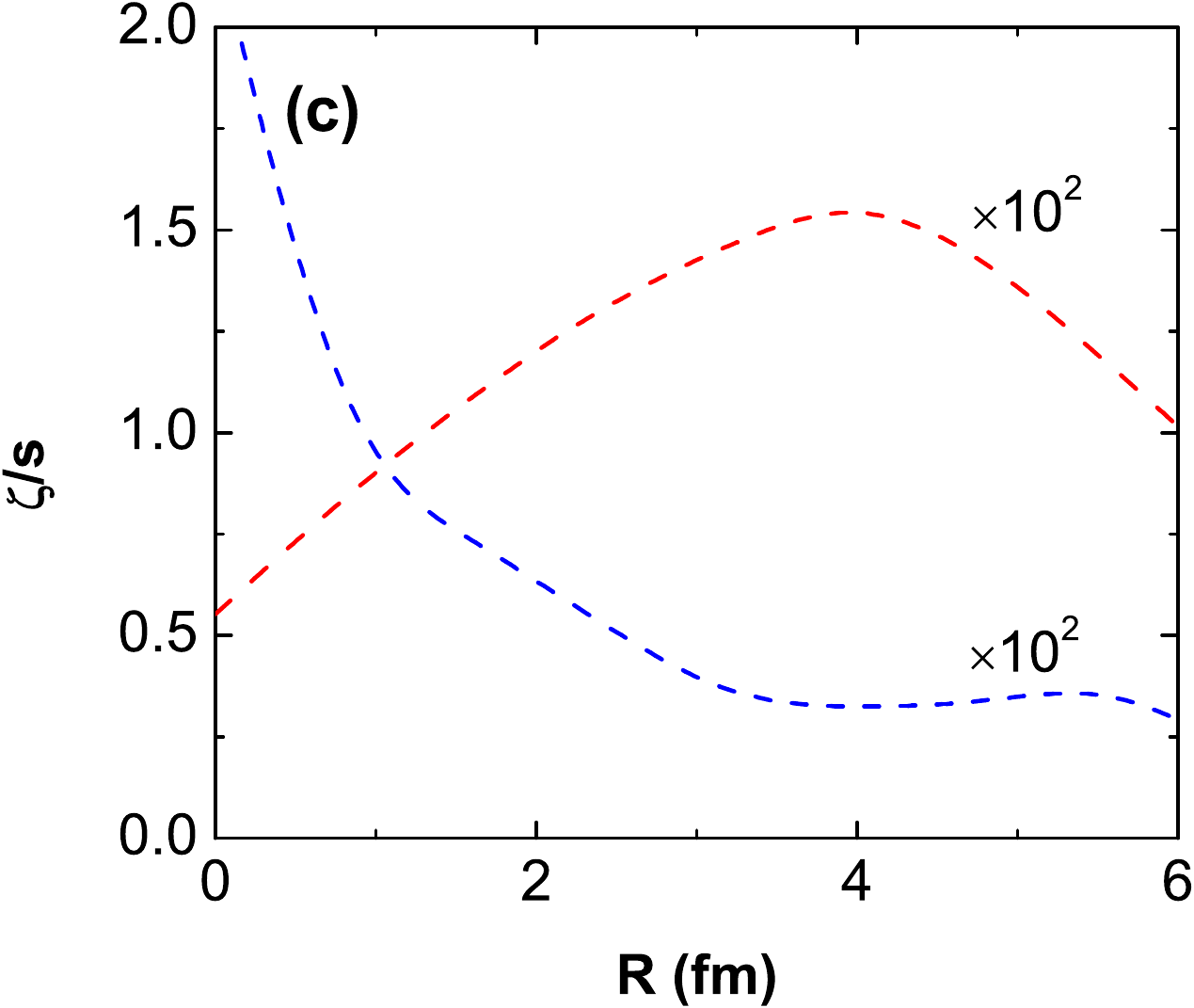}
\caption{(Color online)
Transport and thermodynamic coefficients, (a) speed of sound squared $c_{s}^{2}$, (b) shear viscosity $\eta/s$, and (c) bulk viscosity $\zeta/s$, as functions of the radius.
}
\label{fig2}
\end{center}
\end{figure*}

To understand the properties of medium in collisions, one needs to investigate the transport and thermodynamic properties of system. We re-extracted the effective transport and thermodynamic coefficients, namely, the speed of sound squared ($c_{s}^{2}$), shear viscosity ($\eta/s$), and bulk viscosity ($\zeta/s$). These coefficients as functions of radius, as shown in \Fig{fig2}. By comparing the results of $R_{I}$ and $R_{II}$, one observed that the transport and thermodynamic coefficients are significantly different in the two sets, except for $c_{s}^{2}$ at larger radius ($R\geq 5$ fm). The qualitatively information of \Fig{fig2} is that the $c_{s}^{2}$ decreases smoothly with increasing radius (the behavior of $c_{s}^{2}$ decreases with increasing system volume is also shown in Chiral Mean Field (CMF) model~\cite{Pal:2024eof}), while $\eta/s$, and $\zeta/s$ are significantly different depend on radius in the two sets, respectively.  In the figures: (a) $c_{s}^{2}$ linear dependent on radius is understood as the rapid thermal decomposition of the medium during the expansion process.  (b) A constant shear viscosity for a small radius, and decreases with $R>4$ fm in set I, while the shear viscosity is non-monotonic in set II (the peak close to $R\approx 1$ fm, means that the highest multiplicity is generated in radius $R\leq 1$ fm$^{3}$, similar to ref.~\cite{Silva:2023pei}). (c) The bulk viscosity of the two sets exhibits opposite radius dependence due to the different statistical methods (\eq{rad:01} and \eq{rad:02}) with volume fluctuations. The results underscore the necessity of studying volume-dependent coefficients for accurately quantifying volume fluctuations, as these fluctuations provide the most sensitive probes of the initial geometric fluctuations.

\begin{figure*}[tp]
\begin{center}
\includegraphics[width=0.30\textwidth]{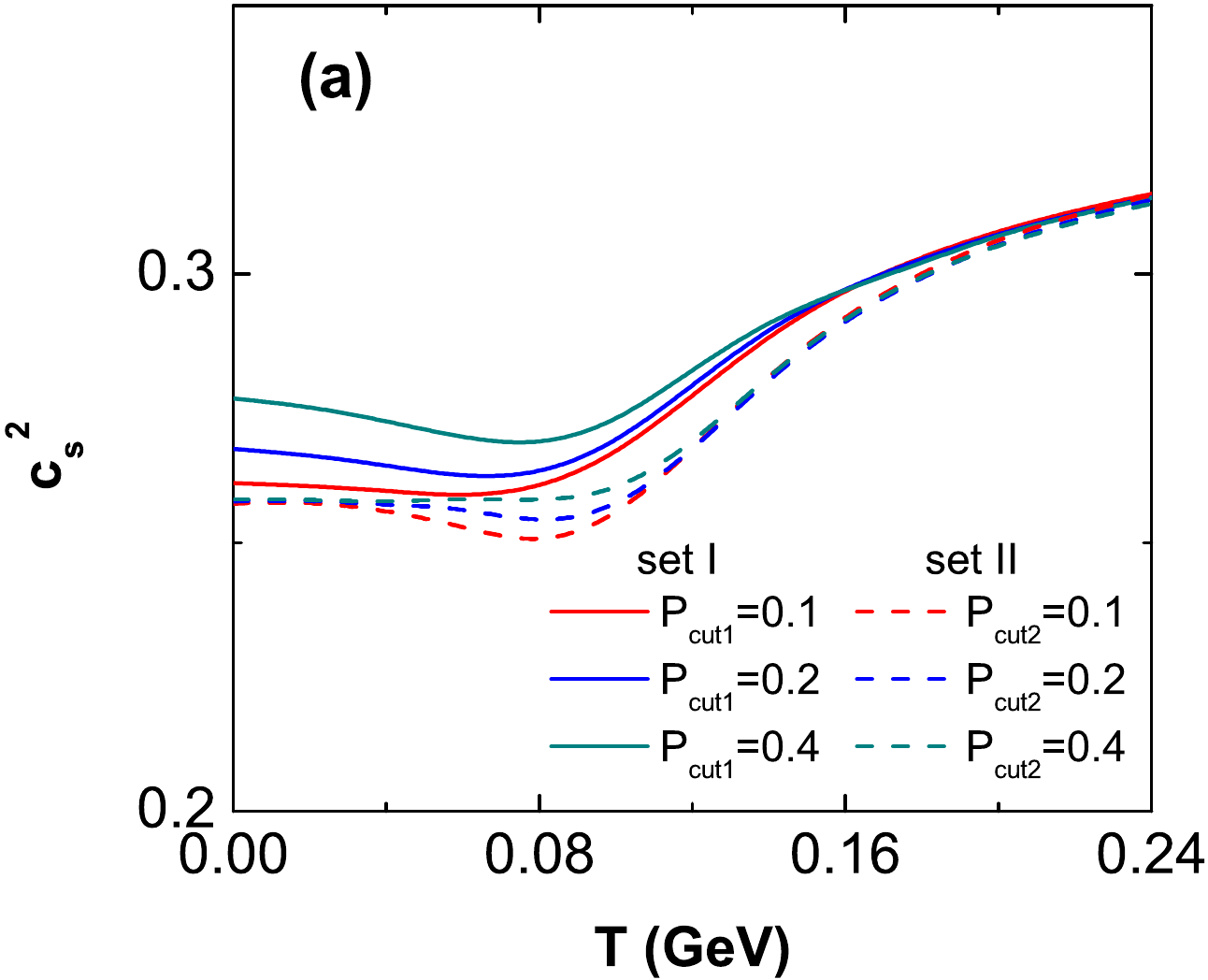}
\hspace{0.10cm}
\includegraphics[width=0.30\textwidth]{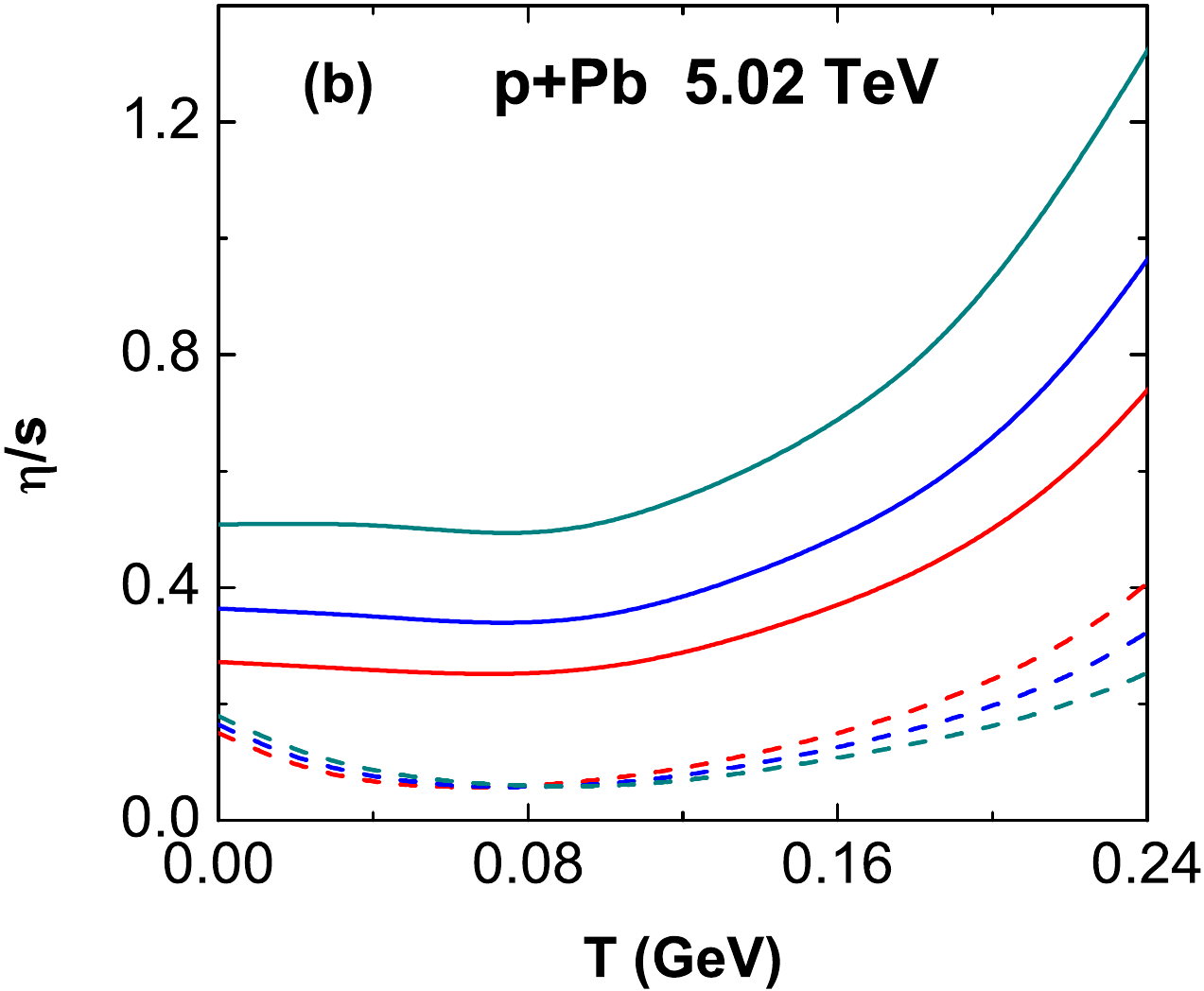}
\hspace{0.10cm}
\includegraphics[width=0.30\textwidth]{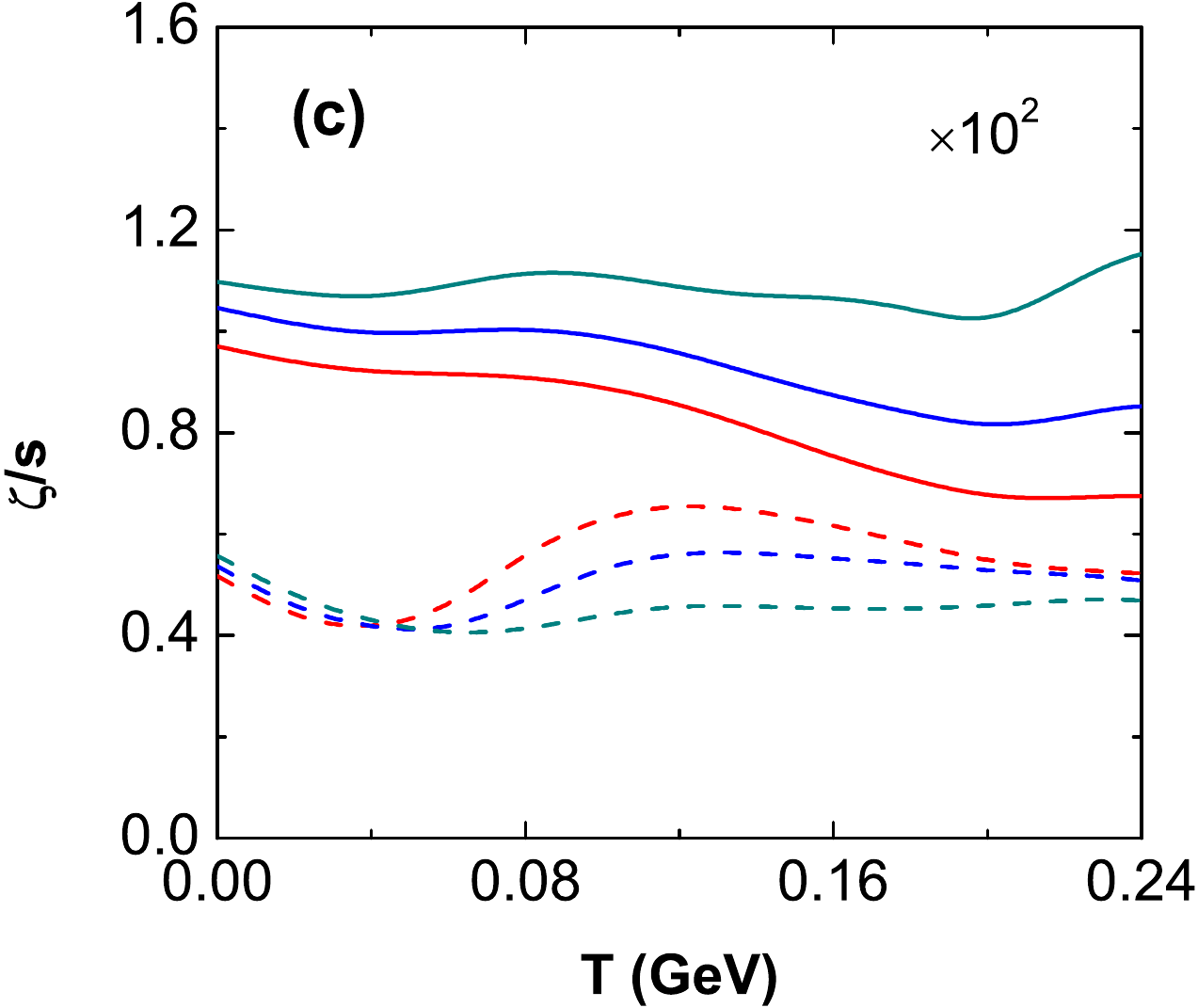}
\caption{(Color online)
Transport and thermodynamic coefficients, (a) speed of sound squared $c_{s}^{2}$, (b) shear viscosity $\eta/s$, and (c) bulk viscosity $\zeta/s$, as functions of local temperature.
}
\label{fig3}
\end{center}\end{figure*}

In \Fig{fig3}, we show the transport and thermodynamic coefficients, such as the $c_{s}^{2}$, $\eta/s$, and $\zeta/s$, as functions of local temperature. These coefficients are compared between the results of sets I and II. 
To investigate the contribution of volume effects, we consider the cut-off methods in the two sets, i.g., the volume is cut-off if the parameter satisfy $P_{cut}\geq 1-\frac{R}{R_{max}}$. In other words, it selects higher multiplicity of events in set I, while it selects higher density energy of events in set II. By taking the value of the volume cut-off parameter from $P_ {cut}=0.1$ to $P_ {cut}=0.4$, one can compare the results between the sets I and II.
Interestingly, we found that the $T$-dependent of coefficients are scaled with the cut-off parameter $P_{cut1}$ in set I, while it is non-monotonically shown with the cut-off parameter $P_{cut2}$ in set II. The transform temperature is close to 0.08 GeV in set II. It should be noted that the re-extraction information on the transport properties of collisions during the evolution of  parton systems, while such partons undergo hadronization as the system temperature decreases to the chemical freeze-out temperature during the evolution process. Therefore, $T$=0.08 GeV should be the chemical freeze-out temperature for the partons that achieve hadronization.

The value of coefficients, are larger in set I than in set II. It can be understood that the transport properties are proportional to the shear and bulk relaxation times in the strong-coupling picture. The initial state is highly non-uniform in small collisions, leading to large non-equilibrium shear and bulk pressures at early times. The shear and bulk viscosities determine the speed at which the system alleviates initial pressure and develops flow. On the other hand, a smaller relaxation time corresponds to stronger local transport processes, as a result, $\eta/s$ and $\zeta/s$ are smaller.

The behavior of the speed of sound squared ($c_{s}^{2}$) is shown in \Fig{fig3} (a). It can be observed that the dip-and-bump effect in temperature ranges and the transform temperature $T_{t}\sim $0.08 GeV. This dip-and-bump behavior is similar to other phenomenological model~\cite{He:2022sos}, and goes accordingly with those reported from the Lattice QCD 2+1 flavor staggered fermion actions p4 and asqtad from~\cite{Bazavov:2009eos}, while they note the critical temperature at $T_{c}\sim $0.16 GeV. Such dip-and-bump behavior gradually disappears has been studied with finite size effects in NJL~\cite{Bhattacharyya:2013tpo}, PNJL~\cite{Zhao:2020tpa} and CMF model~\cite{Pal:2024eof}. This behavior in this work is understood as an increment of effective free degrees of freedom due to partons to achieve hadronization where the temperature is changed from a higher to lower and the transform temperature is close to $T_{t}$=0.08 GeV.

The result of the shear viscosity ($\eta/s$) is shown in \Fig{fig3} (b). One observed that $\eta/s$ rises monotonically with temperature increase for $T\geq$0.08 GeV both in sets I and II, similarly monotonically as shown in Refs.~\cite{Saha:2018tci,Ghosh:2019trf}. The monotonically results are understood as $\eta/s$ decrease with partons undergo hadronization when temperature is falling. $\eta/s$ is a constant value in set I, and it decreases with increasing temperature in set II for $T<$0.08 GeV. $\eta/s$ non-monotonic dependent on temperature is also shown in linear sigma model~\cite{Ghosh:2022svo}. A weak temperature dependence in the temperature range 0.08$\sim T/GeV\sim $0.24 for $P_{cut}$ = 0.4 in set II, may indicate that the QGP is strongly coupled not only in the immediate vicinity of $T_{c}$, but also in a broader temperature region~\cite{Auvinen:2020tdo}.

The result of the bulk viscosity ($\zeta/s$) is shown in \Fig{fig3} (c). One observed that $\zeta/s$ is weakly dependent on temperature in set I while it is concave-and-convex dependent on temperature in set II. In addition, $\zeta/s$ decreases roughly with increasing temperature in the $P_{cut1}$ case, while non-monotonically dependent on temperature in $P_{cut2}$ case. It should be noted that the calculation of bulk viscosity $\zeta/s$ has contributions both from the speed of sound squared $c_{s}^{2}$ and shear viscosity $\eta/s$, and later two quantities are nonlinearly dependent on temperature, as a result, the bulk viscosity is concave-and-convex dependent on temperature. In deeply, the concave-and-convex behavior may be due to the huge volume fluctuations~\cite{Garcia:2023von}.

\begin{figure*}[tp]
\begin{center}
\includegraphics[width=0.40\textwidth]{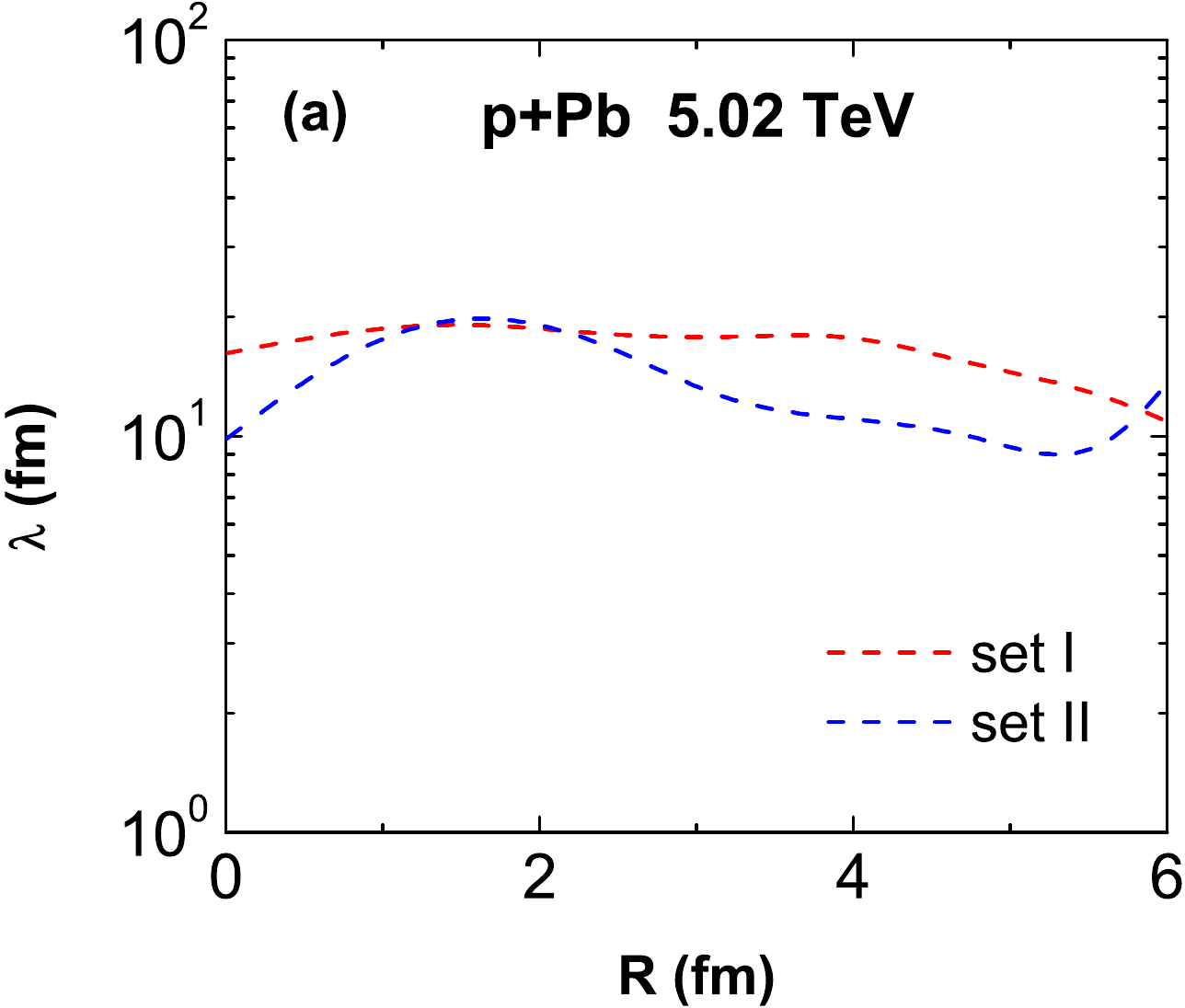}
\hspace{0.10cm}
\includegraphics[width=0.41\textwidth]{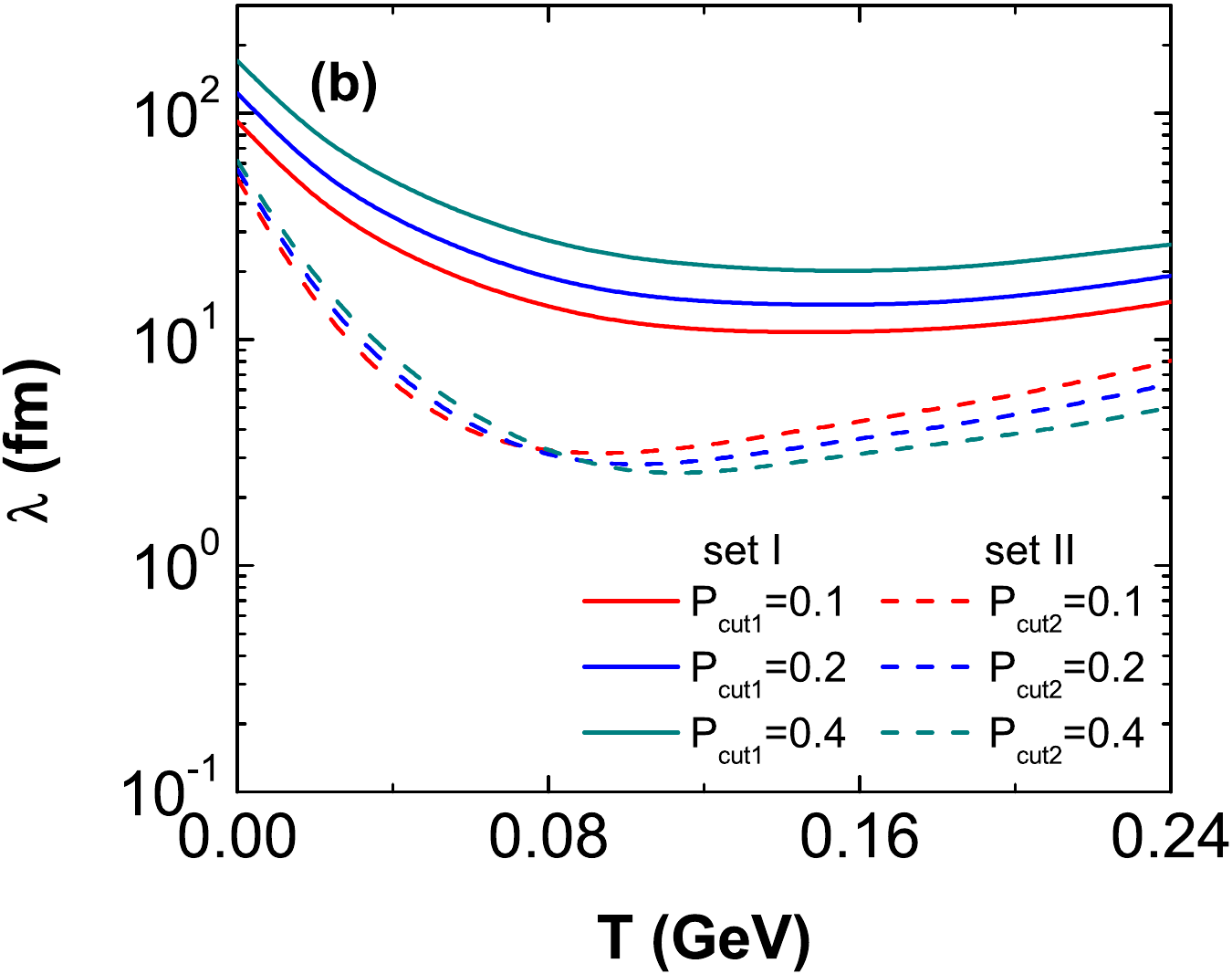}
\caption{(Color online)
The mean free path as functions both of radius and local temperature.
}
\label{fig4}
\end{center}
\end{figure*}

The mean free path ($\lambda$) plays a pivotal role in the characteristic size of a system.
In \Fig{fig4}, the mean free path $\lambda$ as functions both of radius and local temperature. 
As illustrated in \Fig{fig4}, $\lambda$ are significantly different in the sets I and II. This show that $\lambda$ is roughly independent of radius in set I, while its non-monotonically dependent on radius in set II. Such non-monotonic behavior is due to $\lambda$ acquires relevance in calculation of $\eta/s$.

To investigate the temperature dependence of the mean free path, we also considered cut-off parameters in sets I and II. The cut-off results are shown in \Fig{fig4} (b).
In the figure: The mean free path is scaled with the cut-off parameters in set I, while it is non-monotonically influenced by the cut-off parameters and has a minimum crossover temperature of 0.08 GeV in set II. These behaviors can be understood as $\eta/s$-correlated calculations.
At lower temperature, $\lambda$ is higher and it goes on decreases with temperature increasing both in sets I and II, similar results are shown in the String Percolation Model~\cite{Garcia:2023von}.
At higher temperature, $\lambda$ is constant in set I (similar results are shown in Ref.~\cite{Garcia:2023von}), while increased weakly with increasing temperature in set II.
We know that the energy density of the system formed in high-energy collisions increases with increasing temperature. The inflected behavior of temperature dependent of $\lambda$ reflected the mean free path depends on the local energy density. On the other hand, the mean free path affected by the fluctuating volume.

\section{Summary}
\label{sec:sum}

In this work, we explored the transport and thermodynamic coefficients in event-by-event p-Pb collisions at $\sqrt{s_{NN}}$=5.02 TeV using AMPT model.
These transport and thermodynamic coefficients, such as the pressure coefficient $\beta$, speed of sound squared $c_{s}^{2}$, shear viscosity$\eta/s$, bulk viscosity $\zeta/s$ and mean free path $\lambda$ are calculated with kinetic theory approach. 
To study these effective properties, two different definitions of radius have been introduced in the calculation: 
One is weigh of event-by-event charged multiplicity (noted as $R_{I}$ for set I), and the other one is weigh of energy density in a single event (noted as $R_{II}$ for set II).
By comparing the set I and set II:

(a) For the $R$-range, some quantities are monotonically dependent on radius, such as $\beta$ and $c_{s}^{2}$, while
others non-monotonically dependent on radius, such as $\eta/s$, $\zeta/s$ and $\lambda$. The different $R$-range behaviors may imply that the transport and thermodynamic coefficients are significantly affected by the volume fluctuations.

(b) For the $T$-range, transport and thermodynamic coefficients ($c_{s}^{2}$, $\eta/s$, $\zeta/s$ and $\lambda$) are scaled with the cut-off parameters in set I, while it increases with increasing cut-off parameters and decrease with increasing cut-off parameters in set II (the transform temperature is $T_{t}\sim$0.08 GeV). Furthermore, non-monotonically $T$-dependent of transport and thermodynamic coefficients reveal that volume effects and its fluctuations are crucial for understanding the initial fluctuations. 

(c) It must be noted that the kinetic theory cannot completely suppress the non-flow effects.  The impact of non-flow effects on transport and thermodynamic coefficients cannot be ignored in the event-by-event statistical process. In deeply, it requires more comprehensive technology to separate these impacts. It will continue to be discussed in our future work.

The study may provide important information for understanding realtime volume fluctuations in small collisions, while it still requires careful verification.

\section*{Acknowledgements}
This work are supported by the National Natural Science Foundation of China Grant No.~12105057, the National Natural Science Foundation of Guangxi (China) Grant No.~2023GXNSFAA026020,
the Youth Program of Natural Science Foundation of Guangxi (China) Grant No.~2019GXNSFBA245080.

\end{document}